\documentclass[aps,prd,onecolumn,groupedaddress,showpacs,nofootinbib,amssymb]{revtex4-2}
\usepackage[dvips]{graphicx}
\usepackage{amssymb}
\usepackage{amsmath}
\usepackage{graphicx,,color}
\usepackage{amsfonts}
\usepackage{bm}
\usepackage{cancel}
\usepackage{comment}
\usepackage{floatflt}

\newcommand\be{\begin{equation}}
\newcommand\ee{\end{equation}}

\allowdisplaybreaks[4]

\begin{document}

\title{Amplification of the Primordial Gravitational Waves Energy Spectrum by a Kinetic Scalar in $F(R)$ Gravity}
\author{V.K. Oikonomou,$^{1}$}\email{voikonomou@gapps.auth.gr;v.k.oikonomou1979@gmail.com,voikonomou@auth.gr}
\affiliation{$^{1)}$Department of Physics, Aristotle University of
Thessaloniki, Thessaloniki 54124, Greece}


 \tolerance=5000

\begin{abstract}
In this work we consider a combined theoretical framework
comprised by $F(R)$ gravity and a kinetic scalar field. The
kinetic energy of the scalar field dominates over its potential
for all cosmic times, and the kinetic scalar potential is chosen
to be small and non-trivial. In this case, we show that the
primordial gravitational wave energy spectrum of vacuum $F(R)$
gravity is significantly enhanced and can be detectable in future
interferometers. The kinetic scalar thus affects significantly the
inflationary era, since it extends its duration, but also has an
overall amplifying effect on the energy spectrum of pure $F(R)$
gravity primordial gravitational waves. The form of the signal is
characteristic for all these theories, since it is basically flat
and should be detectable from all future gravitational wave
experiments for a wide range of frequencies, unless some unknown
damping factor occurs due to some unknown physical process.
\end{abstract}


\maketitle

\section{Introduction}

The current scientific interest in both astrophysics and
theoretical cosmology is focused on the sky. During the next
fifteen years several eagerly anticipated experiments will
commence giving their first observational data, like the stage 4
Cosmic Microwave Background (CMB) experiments
\cite{CMB-S4:2016ple,SimonsObservatory:2019qwx}, and the
gravitational waves experiments like LISA, DECIGO, Einstein
Telescope and so on
\cite{Hild:2010id,Baker:2019nia,Smith:2019wny,Crowder:2005nr,Smith:2016jqs,Seto:2001qf,Kawamura:2020pcg,Bull:2018lat,LISACosmologyWorkingGroup:2022jok}.
The main focus in all these experiments is to seek for the seeds
of inflation. Indeed, the stage 4 CMB experiments will seek for
the $B$-modes (curl) directly in the CMB polarization, whereas the
gravitational waves experiments will probe directly the
inflationary and post-inflationary epoch, seeking for a stochastic
primordial gravitational wave signal in frequencies for which the
CMB modes grow non-linear. Both the scenarios are exciting and
well anticipated from theoretical cosmologists and also from
theoretical high energy physicists. This is due to the fact that
the future gravitational waves experiments will not only probe the
inflationary era, but also the post-inflationary era, like the
reheating and early radiation epoch. A lot of unknown physics lies
in this era, and therefore with the future gravitational waves
experiments we will have a direct grasp on this mysterious epoch.
To date, the CMB offers insights for primordial inflationary modes
which re-entered the Hubble horizon during the recombination era,
thus these are basically large wavelength modes. Trying to probe
modes with wavelengths smaller than 10$\,$Mpc with the CMB is
futile, since the CMB modes grow non-linear for such small
wavelengths. Thus the future gravitational waves experiments offer
a unique opportunity to probe primordial epochs of the Universe
without resorting to ground experiments which cost more and are
expensive to maintain, without offering any new insights during
the last ten years at least. A signal of a primordial stochastic
tensor perturbation background will be a smoking gun for
inflation, with the latter being of profound theoretical
importance, since it solves many shortcomings of the standard Big
Bang theory, see Refs.
\cite{inflation1,inflation2,inflation3,inflation4} for important
reviews and articles on inflation. The inflationary era can be
described by single scalar field theories, but also in the context
of modified gravity
\cite{reviews1,reviews2,reviews3,reviews4,reviews5}. The advantage
of modified gravity theories over the single scalar field theories
is that in the former theories, one does not need to worry about
the inflaton couplings with the standard model particles in order
to generate the reheating era, which is realized by curvature
fluctuations post-inflationary. Also modified gravity has a
tremendous advantage over the general relativistic descriptions of
the dark energy era, since in the latter case, one needs phantom
scalar fields in order to explain the slightly phantom dark energy
era, allowed by the Planck data. The most important modified
gravity is $F(R)$ gravity
\cite{Nojiri:2003ft,Capozziello:2005ku,Hwang:2001pu,Cognola:2005de,Song:2006ej,Faulkner:2006ub,Olmo:2006eh,Sawicki:2007tf,Faraoni:2007yn,Carloni:2007yv,
Nojiri:2007as,Deruelle:2007pt,Appleby:2008tv,Dunsby:2010wg}, in
the context of which it is possible to even describe inflation and
the dark energy era in a unified manner. The problem with both
$F(R)$ gravity and the scalar field description of the
inflationary era is that these theories produce a quite small
$h^2$-scaled energy spectrum of primordial gravitational waves,
thus the detection of a signal in some future gravitational wave
experiment might directly exclude these theories, in principle
though. In this work we aim to show that a combined theoretical
framework comprised by an $F(R)$ gravity in the presence of a
canonical scalar field which has a large kinetic energy compared
to its scalar potential, can lead to an amplification of the
energy spectrum of the primordial gravitational waves, to a
detectable level. Both scalar fields and higher curvature terms
are motivated by string theory and quantum gravity effects, so in
this work we shall consider their combined effects on the
inflationary era and on the primordial gravitational waves energy
spectrum.

This paper is organized as follows: In section II, we present in
detail the combined $F(R)$ gravity theory in the presence of a
kinetic scalar. We discuss the effects of the kinetic scalar on
the inflationary era and we indicate which is the dominant term
that drives inflation, and we show how the kinetic scalar affects
the duration of the inflationary era and thus can indirectly
affect the reheating temperature. We also calculate the
observational indices of inflation for various reheating
temperatures. In section III, we demonstrate how the current
theory may produce a detectable energy spectrum of the primordial
gravitational waves. Finally the conclusions along with a
discussion follow at the end of the paper.

\section{Inflation in Kinetic Scalar $F(R)$ Gravity}

We shall consider an $F(R)$ gravity in the presence of a canonical
minimally coupled scalar field with the following gravitational
action,
\begin{equation}
\label{mainaction} \mathcal{S}=\int d^4x\sqrt{-g}\left[
\frac{1}{2\kappa^2}F(R)-\frac{1}{2}\partial^{\mu}\phi\partial_{\mu}\phi-V(\phi)+\mathcal{L}_m
\right]\, ,
\end{equation}
with $\kappa^2=\frac{1}{8\pi G}=\frac{1}{M_p^2}$, and $G$ denotes
Newton's gravitational constant, while $M_p$ indicates the reduced
Planck mass. Finally, $\mathcal{L}_m$ represents the Lagrangian
density of the perfect matter fluids present. We shall assume that
both dark matter and radiation perfect fluids are present, in
conjunction with the kinetic scalar and the $F(R)$ gravity. The
$F(R)$ gravity will be considered to have the following form,
\begin{equation}\label{starobinsky}
F(R)=R+\frac{1}{M^2}R^2+R\,e^{\frac{3\Lambda}{R}-\frac{\left(\frac{\Lambda
}{M_p^2}\right)^{1.3} R}{\Lambda }}\, ,
\end{equation}
where $\Lambda$ in Eq. (\ref{starobinsky}) is the present day
cosmological constant, and the parameter $M$ has the form $M=
1.5\times 10^{-5}\left(\frac{N}{50}\right)^{-1}M_p$
\cite{Appleby:2009uf}, where $N$ is the $e$-foldings number.
Assuming that the background metric has the form,
\begin{equation}
\label{metricfrw} ds^2 = - dt^2 + a(t)^2 \sum_{i=1,2,3}
\left(dx^i\right)^2\, ,
\end{equation}
which is a flat Friedmann-Robertson-Walker (FRW) metric, the field
equations for the $F(R)$ gravity scalar field theory take the
following form,
\begin{align}\label{eqnsofmkotion}
& 3 H^2F_R=\frac{RF_R-F}{2}-3H\dot{F}_R+\kappa^2\left(
\rho_r+\rho_m+\frac{1}{2}\dot{\phi}^2+V(\phi)\right)\, ,\\ \notag
& -2\dot{H}F=\kappa^2\dot{\phi}^2+\ddot{F}_R-H\dot{F}_R
+\frac{4\kappa^2}{3}\rho_r\, ,
\end{align}
\begin{equation}\label{scalareqnofmotion}
\ddot{\phi}+3H\dot{\phi}+V'(\phi)=0
\end{equation}
with $F_R=\frac{\partial F}{\partial R}$. Now let us discuss
briefly the kinetic scalar sector of the theory. We basically
require that the kinetic energy of the scalar field always
dominates over its potential, during inflation and
post-inflationary, so basically,
\begin{equation}\label{kineticconstraint}
\dot{\phi}^2\gg V\, ,
\end{equation}
and also we assume the following form of the potential,
\begin{equation}\label{kineticscalarpotential}
V(\phi)=V_0\,M_p^4\,e^{-\delta\, \frac{\phi}{M_p}}\, ,
\end{equation}
where for phenomenological reasoning we shall take $V_0=78.5\times
10^3\times \frac{\Lambda}{H_I^2}$ and $\delta=0.1$, with $H_I$
being the scale of inflation $H_I=10^{13}$GeV, and
$\Lambda=10^{-67}$eV$^2$. As it can be seen, the dimensionless
parameter $V_0$ takes quite small values, thus the kinetic energy
of the scalar field always dominates over its potential during
inflation and post-inflationary. Thus the scalar field is a
kination scalar field, and its field equation is approximately,
\begin{equation}\label{eqnofmotionaxionkinetic}
\ddot{\phi}+3 H\dot{\phi}\simeq 0\, ,
\end{equation}
which yields,
\begin{equation}\label{axionkineticfieldeqn}
\dot{\phi}\sim a^{-3}\, ,
\end{equation}
therefore the energy density of the kinetic scalar, which is
defined as, $\rho_{\phi}=\frac{\dot{\phi}^2}{2}+V(\phi)\simeq
\frac{\dot{\phi}^2}{2}$, redshifts during inflation and
post-inflationary as $\rho_{\phi}\sim a^{-6}$. Hence the kinetic
scalar has a stiff equation of state (EoS) parameter
$\omega_{\phi}=1$, and behaves essentially as a stiff fluid during
and after the inflationary era. Now, due to this fact, the scalar
field contribution during inflation can be disregarded for the
same reason that the dark matter and radiation perfect fluids are
disregarded during inflation. Also among the two terms in the
$F(R)$ gravity functional form in Eq. (\ref{starobinsky}), only
the $R^2$ term dominates at early times, thus it controls the
inflationary era. Thus at early times we have,
\begin{equation}\label{effectivelagrangian2}
F(R)\simeq R+\frac{1}{M^2}R^2\, .
\end{equation}
hence the field equations become,
\begin{equation}\label{patsunappendix}
\ddot{H}-\frac{\dot{H}^2}{2H}+\frac{H\,M^2}{2}=-3H\dot{H}\, .
\end{equation}
and by imposing the slow-roll conditions,
\begin{equation}\label{patsunappendix1}
-\frac{M^2}{6}=\dot{H}\, ,
\end{equation}
we get a quasi-de Sitter evolution,
\begin{equation}\label{quasidesitter}
H(t)=H_I-\frac{M^2}{6} t\, ,
\end{equation}
with $H_I$ being the scale of inflation. Hence at the level of the
background evolution, which stems from the field equations, the
$R^2$ gravity controls the dynamics, therefore one could claim
that the kinetic scalar field does not affect the inflationary
era. This is not true however at the cosmological perturbations
level, since the scalar field affects the slow-roll indices and
eventually may affect the observational indices. Now, for the
$f(R,\phi)$ theory we consider in this paper, the slow-roll
indices are defined as
\cite{Hwang:2005hb,reviews1,Odintsov:2020thl},
\begin{equation}
\label{restofparametersfr}\epsilon_1=-\frac{\dot{H}}{H^2}, \quad
\epsilon_2=\frac{\ddot{\phi}}{H\dot{\phi}}\, ,\quad \epsilon_3=
\frac{\dot{F}_R}{2HF_R}\, ,\quad
\epsilon_4=\frac{\dot{E}}{2H\,E}\,
 ,
\end{equation}
with $E$ being,
\begin{equation}\label{eparameter}
E=F_R+\frac{3\dot{F}_R^2}{2\kappa^2\dot{\phi}^2}\, .
\end{equation}
Due to the fact that the scalar field is kinetic energy dominated,
from its equation of motion we get that $\epsilon_2=-3$, hence the
scalar field obeys a constant-roll evolutions. This parameter
eventually enters in the final expression of the slow-roll
parameter $\epsilon_4$, and as it shown in detail in Ref.
\cite{submitted}, we get,
\begin{equation}\label{epsilon4finalnew}
\epsilon_4\simeq -\frac{x}{2}\epsilon_1-\epsilon_1-\epsilon_2\, ,
\end{equation}
with $x$ being defined as follows,
\begin{equation}\label{parameterx}
x=\frac{48 F_{RRR}H^2}{F_{RR}}\, .
\end{equation}
The spectral index of the scalar curvature fluctuations for the
$f(R,\phi)$ gravity at hand, has the following form
\cite{Hwang:2005hb,reviews1,Odintsov:2020thl},
\begin{equation}\label{scalarspectralindex}
n_{\mathcal{S}}=1-4\epsilon_1-2\epsilon_2+2\epsilon_3-2\epsilon_4\,
,
\end{equation}
hence remarkably,  the slow-roll index $\epsilon_2$ is eliminated
from the final expression, hence the spectral index becomes,
\begin{equation}\label{spectralindexfinalform}
n_{\mathcal{S}}\simeq 1-(2-x)\epsilon_1+2\epsilon_3\, .
\end{equation}
Furthermore, the tensor-to-scalar ratio for the current theory has
the following form \cite{Hwang:2005hb,reviews1,Odintsov:2020thl},
\begin{equation}\label{tensortoscaalrratio}
r\simeq 48\epsilon_1^2\, .
\end{equation}
In effect, the observational indices take the form
$n_{\mathcal{S}}\sim 1-\frac{2}{N}$ and $r\sim \frac{12}{N^2}$,
and also the tensor spectral index is $n_T=-\frac{1}{2N^2}$, thus
these are identical to those corresponding to the $R^2$ model.
Remarkably, the kinetic scalar field does not affect the dynamics
of inflation even at the cosmological perturbations level, however
it affects the duration of the inflationary era. The mechanism is
simple, as inflation proceeds and the cosmological system reaches
its final unstable quasi-de Sitter attractor, the curvature
fluctuations become strong and the scalar field energy density
starts to dominate the evolution before the other two perfect
fluids start to control the evolution. Thus the Universe
experiences a short period of kination, before it enters the
reheating era. The complete analysis of this mechanism is better
explained in Ref. \cite{submitted}. Thus, since the total EoS is
$w=1$, the total number of the $e$-foldings by including the short
stiff era is given by
\cite{Adshead:2010mc,Munoz:2014eqa,Liddle:2003as},
\begin{equation}\label{efoldingsmainrelation}
N=56.12-\ln \left( \frac{k}{k_*}\right)+\frac{1}{3(1+w)}\ln \left(
\frac{2}{3}\right)+\ln \left(
\frac{\rho_k^{1/4}}{\rho_{end}^{1/4}}\right)+\frac{1-3w}{3(1+w)}\ln
\left( \frac{\rho_{reh}^{1/4}}{\rho_{end}^{1/4}}\right)+\ln \left(
\frac{\rho_k^{1/4}}{10^{16}\mathrm{GeV}}\right)\, ,
\end{equation}
where $\rho_k$ stands for the total energy density of the Universe
at first horizon crossing of the mode $k$, $\rho_{end}$ is the
energy density of the Universe when inflation ends eventually, and
$\rho_{reh}$ is the energy density of the Universe at the end of
the reheating era, corresponding to a reheating temperature $T_R$.
The above expression for the $e$-foldings number can be expressed
in terms of the temperatures at the various epochs by using
$\rho=\frac{\pi^2}{30}g_*T^4$, assuming a constant number of
particles during and after the inflationary era. Also we shall
take the pivot scale to be $k_*=0.05$Mpc$^{-1}$. We can directly
find the total number of $e$-foldings and hence the corresponding
values of the observational indices for various reheating
temperatures. We shall assume three distinct reheating
temperatures, namely $T_R=10^{12}$GeV, $T_R=10^{7}$GeV
$T_R=10^{2}$GeV. Having low reheating temperatures can be easily
understood in the context of this kination dominated epoch, since
the reheating era starts at a later time, compared with the vacuum
$R^2$ model. Apart from this line of reasoning, there exist other
phenomenological reasons that can explain a low-reheating
temperature, see for example \cite{Hasegawa:2019jsa}. Now let us
compute the observational indices for the three distinct reheating
temperatures, because we shall need these values for the
calculation of the energy spectrum of the primordial gravitational
waves. For the reheating temperature  $T_R=10^{12}$GeV, the
$e$-foldings number reads, $N=65.41$ and the observational indices
in this case read, $n_{\mathcal{S}}=0.969424$, $r=0.00280474$,
$n_T=-0.000116864$ while for $T_R=10^{7}$GeV, the $e$-foldings
number reads, $N=70.7842$ and the observational indices in this
case read, $n_{\mathcal{S}}=0.971745$, $r=0.00239502$,
$n_T=-0.00009979$. Finally, for $T_R=10^{2}$GeV, the $e$-foldings
number reads, $N=74.621$ and the observational indices in this
case read, $n_{\mathcal{S}}=0.9717$, $r=0.0021$, $n_T=-0.000089$.
We shall use these values in the next section, where we shall
calculate the energy spectrum of the primordial gravitational
waves.

\section{Energy Spectrum of Primordial Gravitational Waves for the Kinetic Scalar $F(R)$ Gravity}

Having described the evolution of the combined $F(R)$ gravity
kinetic scalar model, in this section we shall investigate the
effect of this kinetic scalar on the energy spectrum of the
primordial gravitational waves. There exists a vast literature
related to primordial gravitational waves, see Refs.
\cite{Kamionkowski:2015yta,Denissenya:2018mqs,Turner:1993vb,Boyle:2005se,Zhang:2005nw,Schutz:2010xm,Sathyaprakash:2009xs,Caprini:2018mtu,
Arutyunov:2016kve,Kuroyanagi:2008ye,Clarke:2020bil,Kuroyanagi:2014nba,Nakayama:2009ce,Smith:2005mm,Giovannini:2008tm,
Liu:2015psa,Zhao:2013bba,Vagnozzi:2020gtf,Watanabe:2006qe,Kamionkowski:1993fg,Giare:2020vss,Kuroyanagi:2020sfw,Zhao:2006mm,
Nishizawa:2017nef,Arai:2017hxj,Bellini:2014fua,Nunes:2018zot,DAgostino:2019hvh,Mitra:2020vzq,Kuroyanagi:2011fy,Campeti:2020xwn,
Nishizawa:2014zra,Zhao:2006eb,Cheng:2021nyo,Nishizawa:2011eq,Chongchitnan:2006pe,Lasky:2015lej,Guzzetti:2016mkm,Ben-Dayan:2019gll,
Nakayama:2008wy,Capozziello:2017vdi,Capozziello:2008fn,Capozziello:2008rq,Cai:2021uup,Cai:2018dig,Odintsov:2021kup,Benetti:2021uea,Lin:2021vwc,Zhang:2021vak,Odintsov:2021urx,Pritchard:2004qp,Zhang:2005nv,Baskaran:2006qs,Oikonomou:2022xoq,Odintsov:2022cbm,Odintsov:2022sdk,Kawai:2017kqt,Odintsov:2022hxu,Gao:2019liu,Oikonomou:2022yle,Oikonomou:2022pdf,Vagnozzi:2022qmc,Gerbino:2016sgw,Casalino:2018wnc,Casalino:2018tcd,Visinelli:2017bny,Visinelli:2018utg,Kinney:2018nny,Giare:2020plo,Ricciardone:2016ddg}
for an important stream of related articles.

Now let us directly proceed to quantify the effect of the $F(R)$
gravity kinetic scalar theory on the energy spectrum of the
primordial gravitational waves, for details see
\cite{Odintsov:2021kup}. The Fourier transformation of the
primordial tensor perturbation $h_{i j}$ satisfies the following
differential equation,
\begin{equation}\label{mainevolutiondiffeqnfrgravity}
\ddot{h}(k)+\left(3+a_M \right)H\dot{h}(k)+\frac{k^2}{a^2}h(k)=0\,
,
\end{equation}
with the parameter $\alpha_M$ being of significant importance and
it is equal to,
\begin{equation}\label{amfrgravity}
a_M=\frac{\dot{Q}_t}{Q_tH}\, ,
\end{equation}
with the function $Q_t$ is basically determined by the modified
gravity theory at hand. For a review on the various forms of the
function $Q_t$ and for general considerations related to
primordial gravitational waves in modified gravity, see Ref.
\cite{Odintsov:2022cbm}. Since the theoretical framework in our
case is a type of $f(R,\phi)$ gravity of the form,
\begin{equation}\label{action1}
\mathcal{S}=\int
\mathrm{d}^4x\sqrt{-g}\Big{(}\frac{f(R,\phi)}{2}-\frac{1}{2}\partial^{\mu}\phi\partial_{\mu}\phi-V(\phi)\Big{)}\,
,
\end{equation}
for the case at hand, the function $Q_t$ is equal to
$Q_t=\frac{1}{\kappa^2}\frac{\partial f(R,\phi)}{\partial R}$.
Hence, for the case at hand, the parameter $a_M$ is equal to,
\begin{equation}\label{amfrphi}
a_M=\frac{\frac{\partial^2f}{\partial R \partial
\phi}\dot{\phi}+\frac{\partial^2f}{\partial
R^2}\dot{R}}{\frac{\partial f}{\partial R}H}\, .
\end{equation}
In our case, the parameter $a_M$ reduces to,
\begin{equation}\label{amfrgravity}
a_M=\frac{F_{RR}\dot{R}}{F_RH}\, .
\end{equation}
We shall make use of the relation above for the calculation of the
overall effect of modified gravity on the energy spectrum of the
primordial gravitational waves. The differential equation that
governs the evolution of the tensor perturbations expressed in
terms of the conformal time takes the following form,
\begin{equation}\label{mainevolutiondiffeqnfrgravityconftime}
h''(k)+\left(2+a_M \right)\mathcal{H} h'(k)+k^2h(k)=0\, ,
\end{equation}
with the ``prime'' denoting differentiation with respect to $\tau$
and also $\mathcal{H}=\frac{a'}{a}$. The above equation can be
solved by using a WKB method, so assuming a WKB solution of the
form $h_{ij}=\mathcal{A}e^{i\mathcal{B}}h_{ij}^{GR}$
\cite{Nishizawa:2017nef,Arai:2017hxj}, we have,
\begin{equation}\label{mainsolutionwkb}
h=e^{-\mathcal{D}}h_{GR}\, ,
\end{equation}
with $h_{i j}=h e_{i j}$, and note that $h_{GR}$ is the general
relativistic waveform. Note that the WKB solution is valid only
for subhorizon modes during the reheating era. More importantly,
the parameter $\mathcal{D}$ is defined as follows,
\begin{equation}\label{dform}
\mathcal{D}=\frac{1}{2}\int^{\tau}a_M\mathcal{H}{\rm
d}\tau_1=\frac{1}{2}\int_0^z\frac{a_M}{1+z'}{\rm d z'}\, ,
\end{equation}
and the calculation of the parameter $\mathcal{D}$ is the main
focus when studying modified gravity effects on primordial
gravitational waves. This is what we will do in this section for
the combined $F(R)$ gravity kinetic scalar theory. Thus for the
calculation we will need to integrate from present day up to a
high redshift corresponding to the reheating era. For the purposes
of this work we shall integrate up to $z_f=10^{12}$ which goes
deeply in the reheating era.
\begin{figure}[h!]
\centering
\includegraphics[width=40pc]{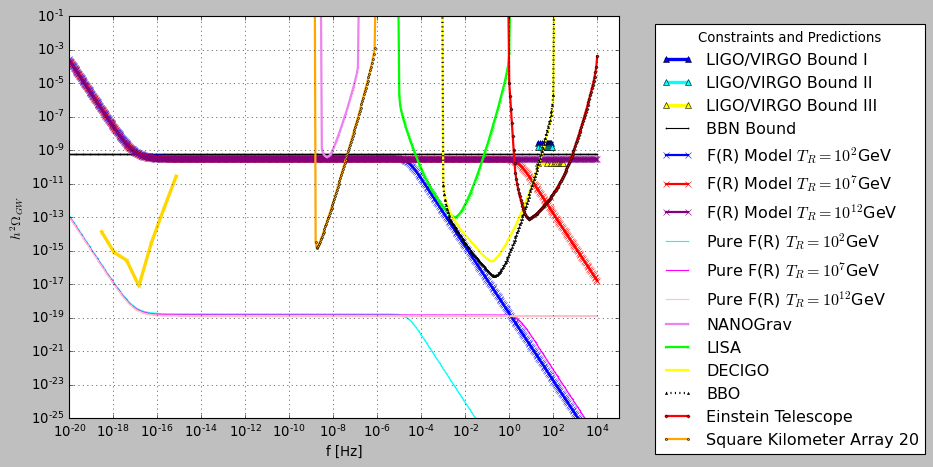}
\caption{The $h^2$-scaled gravitational wave energy spectrum for
the combined $F(R)$ gravity kinetic scalar theory versus the
sensitivity curves of future gravitational waves experiments.
Three distinct reheating temperatures are considered.}
\label{plotfinalfrpure2}
\end{figure}
By including the WKB contribution to the energy spectrum of the
primordial gravitational waves, the latter takes the following
form
\cite{Boyle:2005se,Nishizawa:2017nef,Arai:2017hxj,Nunes:2018zot,Liu:2015psa,Zhao:2013bba,Odintsov:2021kup},
\begin{align}
\label{GWspecfR}
    &\Omega_{\rm gw}(f)=e^{-2\mathcal{D}}\times \frac{k^2}{12H_0^2}r\mathcal{P}_{\zeta}(k_{ref})\left(\frac{k}{k_{ref}}
\right)^{n_T} \left ( \frac{\Omega_m}{\Omega_\Lambda} \right )^2
    \left ( \frac{g_*(T_{\rm in})}{g_{*0}} \right )
    \left ( \frac{g_{*s0}}{g_{*s}(T_{\rm in})} \right )^{4/3} \nonumber  \left (\overline{ \frac{3j_1(k\tau_0)}{k\tau_0} } \right )^2
    T_1^2\left ( x_{\rm eq} \right )
    T_2^2\left ( x_R \right )\, ,
\end{align}
with $k_{ref}=0.002$$\,$Mpc$^{-1}$ being the pivot scale of CMB.
In order to calculate the integral of Eq. (\ref{dform}), we need
to solve numerically the Friedmann equation from redshift zero up
to the redshift $z_f=10^{12}$. We can calculate directly the
integral of Eq. (\ref{dform}), by numerically solving the
Friedmann equation using appropriate initial conditions. This
integration will also reveal the late-time behavior of the
combined $F(R)$ gravity kinetic scalar model, but also the
behavior of the scalar field up to a redshift $z_f=10^{12}$. In
order to do so, we bring the Friedmann equation into an
appropriate form by using the following relations,
\begin{equation}
\centering \dot H=-H(1+z)H'\, ,
\end{equation}
\begin{equation}
\centering \dot\phi=-H(1+z)\phi'\, ,
\end{equation}
\begin{equation}
\centering
\ddot\phi=H^2(1+z)^2\phi''+H^2(1+z)\phi'+HH'(1+z)^2\phi'\, ,
\end{equation}
\begin{equation}
\centering \dot f_R=\dot Rf_{RR}+\dot\phi f_{R\phi}\, ,
\end{equation}
\begin{equation}
\centering \dot
R=6H(1+z)^2\left(HH''+(H')^2-\frac{3HH'}{1+z}\right)\, .
\end{equation}
Also we introduce the statefinder function $y_H(z)$
\cite{Hu:2007nk,Bamba:2012qi,Odintsov:2020qyw},
\begin{equation}
\centering \label{yH} y_H=\frac{\rho_{DE}}{\rho_{d0}}\, ,
\end{equation}
where $\rho_{DE}$ denotes the dark matter energy density and
$\rho_{d0}$ stands for the value of density for non-relativistic
matter at present day. The dark energy density is,
\begin{equation}
\centering \label{DEdensity}
\rho_{DE}=\frac{1}{2}\dot\phi^2+V+\frac{f_R
R-f}{2\kappa^2}-\frac{3H\dot
f_R}{\kappa^2}+\frac{3H^2}{\kappa^2}(1-f_R)\, ,
\end{equation}
and the dark energy pressure is,
\begin{equation}
\centering \label{PDE} P_{DE}=-V-24\dot\xi H^3-8\dot\xi H\dot
H-\frac{f_R R-f}{2\kappa^2}-\frac{2\dot H}{\kappa^2}(1-f_R)\, ,
\end{equation}
with,
\begin{equation}
\centering \label{conteqDE} \dot\rho_{DE}+3H(\rho_{DE}+P_{DE})=0\,
.
\end{equation}
Thus the Friedmann equation takes the form,
\begin{equation}
\centering \label{motion6}
\frac{3H^2}{\kappa^2}=\rho_{(m)}+\rho_{DE}\, ,
\end{equation}
\begin{equation}
\centering \label{motion7} -\frac{2\dot
H}{\kappa^2}=\rho_{(m)}+P_{(m)}+\rho_{DE}+P_{DE}\, ,
\end{equation}
where $\rho_{(m)}$ and $P_{(m)}$ are the total energy density and
total pressure of the perfect matter fluids present. In terms of
the statefinder function $y_H(z)$, the Hubble rate is written as
follows,
\begin{equation}
\centering \label{H}
H^2=m_s^2\left(y_H(z)+\frac{\rho_{(m)}}{\rho_{d0}}\right)\, ,
\end{equation}
with $m_s^2=\kappa^2\frac{\rho_{d0}}{3}=1.87101\cdot10^{-67}$,
while the higher derivatives of the Hubble rate read,
\begin{equation}
\centering \label{H'}
HH'=\frac{m_s^2}{2}\left(y_H'+\frac{\rho_{(m)}'}{\rho_{d0}}\right)\,
,
\end{equation}
\begin{equation}
\centering \label{H''}
H'^2+HH''=\frac{m_s^2}{2}\left(y_H''+\frac{\rho_{(m)}''}{\rho_{d0}}\right)\,
.
\end{equation}
Finally, with regard to the dark energy era, the dark energy
equation of state parameter $\omega_{DE}$ and the dark energy
density parameter $\Omega_{DE}$, they can be expressed in terms of
the statefinder function $y_H$ as follows
\cite{Bamba:2012qi,Odintsov:2020qyw},
\begin{align}
\centering \label{DE}
\omega_{DE}&=-1+\frac{1+z}{3}\frac{d\ln{y_H}}{dz}&\Omega_{DE}&=\frac{y_H}{y_H+\frac{\rho_{(m)}}{\rho_{d0}}}\,
.
\end{align}
Also the initial conditions for the statefinder quantity $y_H(z)$
and for the scalar field $\phi (z)$ are the following,
\begin{equation}\label{initialconditionsforscalarandyH}
y_H(0)=\frac{\Lambda }{3 \text{ms}^2}\,
,\,\,\,y_H'(0)=0,\,\,\,\phi
(0)=10^{-16}\,M_p,\,\,\,\phi'(0)=10^{-17}\,M_p^2\, ,
\end{equation}
so we start basically from a late-time de Sitter era. The energy
density parameter of dark energy at present day is equal to
$\Omega_{DE}(0)=0.679335$ and the corresponding dark energy EoS
parameter $\omega_{DE}(0)=-1$. Upon solving numerically the
Friedmann equation, we obtain the numerical solution for the
statefinder function $y_H(z)$, and thus we can numerically perform
the integration of Eq. (\ref{dform}) and the result of the
integration is $\int_{0}^{z_f}\frac{a_{M}}{1+z'}{\rm d z'}\simeq
-23$ This leads to a large amplification of the pure $F(R)$
gravity signal, and in Fig. \ref{plotfinalfrpure2} we plot the
predicted signal of the combined $F(R)$ gravity kinetic scalar
theory, and also the sensitivity curves of the future
gravitational waves experiments. Note that our WKB solution is
valid for subhorizon modes during reheating and beyond, so the
small frequency part of the plot must be disregarded, for
frequencies smaller that $f<10^{-10}\,$Hz. In the plot of Fig.
\ref{plotfinalfrpure2} we included three distinct reheating
temperatures, a high one $T_R=10^{12}$GeV (purple curve), an
intermediate $T_R=10^{7}$GeV (red curve) and a low reheating
temperature $T_R=10^{2}$GeV (blue curve). From Fig.
\ref{plotfinalfrpure2} it is obvious that the signal is detectable
from all the future gravitational wave experiments, for the high
and intermediate temperatures, except for the low reheating
temperature\footnote{For the sensitivity curves of the future
primordial gravitational wave experiments see
\cite{Breitbach:2018ddu}}. It is also mentionable that the signal
of the theory at hand is somewhat uniform and it leads to a nearly
constant $h^2$-scaled gravitational wave energy spectrum. This
indicates that if the signal corresponding to this class of
theories is detected by one experiment, then it will be detected
by all experiments and will have the same amplitude and energy
spectrum. This feature is of particular importance since it is a
characteristic of this class of theories. We shall discuss further
possibilities on how to discriminate modified gravity theories by
using primordial gravitational waves in the next section. One
thing to mention which is somewhat important is that the
amplification of the primordial gravitational wave energy spectrum
only occurs if the potential is small compared to the kinetic
term, but if someone chooses a potential that does not satisfy
this condition, or if the potential is trivial, no amplification
occurs. The potential must be no trivial and present, but it must
be dominated by the kinetic term of the scalar field.

Before closing, we need to mention that in our analysis we did
include the most recent constraints imposed by the LIGO/VIRGO
collaboration \cite{KAGRA:2021kbb}, which constrain the energy
spectrum of the primordial waves to be $\Omega_{GW}\leq 5.8\times
10^{-9}$ for a flat (frequency independent) gravitational wave
background in the frequency range $(20-76.6)$Hz (LIGO VIRGO BOUND
I), and also $\Omega_{GW}\leq 3.4\times 10^{-9}$ for a power-law
gravitational background with a spectral index of $2/3$ in the
frequency range $(20-90.6)$Hz (LIGO VIRGO BOUND II) and
furthermore $\Omega_{GW}\leq 3.9\times 10^{-10}$ for a spectral
index of $3$, in the frequency range $(20-291.6)$Hz (LIGO VIRGO
BOUND III). Also we included the constraints from the BBN
\cite{Kuroyanagi:2014nba}. As it can be seen in Fig.
\ref{plotfinalfrpure2}, the resulting combined $F(R)$ gravity
kinetic scalar pass the LIGO/VIRGO bounds I and II and also
satisfy the BBN, save the large reheating temperature combined
$F(R)$ gravity kinetic scalar model, which is marginally
compatible with the BBN constraints and completely incompatible
with the LIGO/VIRGO bound III. We also included in our study the
pure $F(R)$ gravity cases for three reheating temperatures, see
the lower lines in Fig. \ref{plotfinalfrpure2}. As it can be seen,
the signal is way lower than the sensitivity curves of all the
future experiments.

We also need to note that the amplification is sensitive to
parameter changes, thus the amplification is model dependent. This
shows that there is freedom in the choices of the parameters,
however it would generally be difficult to pin-point which model
produces a future detected signal. However, with this work we
aimed to show that it is possible to have an enhancement in the
primordial gravitational wave energy spectrum if one combines a
kinetic energy dominated scalar field with $F(R)$ gravity, with
the scalar field being some light axion-like particle. Of course,
more complicated scenarios can occur, for example a complication
may be induced by a prolonged inflationary era. This scenario is
interesting itself, and is worth studying in the future because
the scenario we studied here also leads to a prolonged
inflationary era, slightly prolonged though. In conclusion, many
features of the standard GR compatible scalar field theory may
change when $F(R)$ gravity corrections are taken into account.
This work aimed in pointing out this feature of $f(R,\phi)$
theories.

\section{Conclusions and Discussion}

In this work we studied a combined cosmological framework
comprised by an $F(R)$ gravity and a kinetic canonical scalar
field. The terminology ``kinetic'' for the scalar field is used
because we assumed that the kinetic energy of the scalar field
dominates over its potential for its whole evolution, without
however disregarding its potential. Due to the fact that the
scalar field is kinetic, this means that during inflation, and of
course beyond inflation, the scalar field obeys a constant-roll
evolution and it has a stiff equation of state. However, being a
stiff fluid, and thus redshifting as $a^{-6}$, this means that it
does not affect the inflationary dynamics at the equations of
motion level. Nevertheless, we investigated two more ways that
this kinetic scalar might affect the dynamics of inflation, one
way is at the cosmological perturbations level, via the second
slow-roll parameter, and the other way is via the extension of the
$e$-foldings number for the inflationary era. As we showed, the
effects of the kinetic scalar at the cosmological perturbation
level cancels in the final functional form of the observational
indices and specifically of the spectral index of the scalar
curvature perturbations. On the other hand, the kinetic scalar
affects the duration of inflation. Indeed, when the unstable
quasi-de Sitter attractor of the $F(R)$ gravity model is reached,
the $F(R)$ gravity no longer controls the evolution, thus the
kinetic scalar starts to dominate the evolution. This makes the
total EoS parameter of the Universe to be equal to unity, thus the
Universe passes through a kination era. This feature affects
directly the duration of the inflationary era, and thus the
observational indices.

After discussing inflation in this kinetic scalar $F(R)$ gravity
framework, we investigated the implications of the combined
framework on the primordial gravitational wave energy spectrum.
Apart from the observational indices of inflation, and
specifically the tensor-to-scalar ratio and the tensor spectral
index, which are different for the combined $F(R)$ gravity kinetic
scalar framework compared to the vacuum $F(R)$ gravity case, the
energy spectrum of the former is significantly enhanced compared
to the latter theory. We numerically calculated the amplification
factor for the combined $F(R)$ gravity kinetic scalar framework,
starting from the present day epoch up to redshifts of the order
$10^{12}$ which is deeply in the radiation domination era. As we
showed, the energy spectrum of the primordial gravitational wave
is significantly enhanced and moreover the signal is flat and
appears to be of the same order for a wide range of frequencies. A
notable feature of the calculation is the form of the potential,
and this enhancement seems to work only for potentials that tend
to zero for all scalar field values, such as the exponential
potentials. It is remarkable though that if we switch off the
potential, the enhancement mechanism does not work. One needs a
kinetic scalar field with a non-trivial scalar potential which
goes smoothly to zero for large scalar field values. Quadratic
potentials and in general power-law potentials with positive
values of the exponents do not work. Another important feature of
this framework is the form of the signal. It has the same form as
the vacuum $F(R)$ gravity case, although significantly enhanced.
Thus, this class of theories leads to a signal that will be
detectable from all the future gravitational wave experiments, if
the reheating temperature is large enough though. Hence, if a
signal is detected and the signal is of the same order of
magnitude for all the detectors, the underlying theory is very
likely some sort of combined scalar field $F(R)$ gravity theory,
with the amplification mechanism we showed in this article. If
however the signal is detected by some and not all the detectors,
this would either indicate that the underlying theory is not of
the form we discussed in this paper, or there is a strong damping
mechanism in specific frequency range. Regarding the latter
damping, it is interesting to note there exist various damping
mechanisms, for example supersymmetry breaking during reheating,
or even the chameleon mechanism may lead to a serious suppression
of the energy spectrum, see for example
\cite{Katsuragawa:2019uto}. Thus the future is highly anticipated
by theoretical physicists to answer (hopefully) their deep and
unanswered for decades questions.

\end{document}